\documentstyle[12pt]{article}
\newcommand{\be}{\begin{equation}}
\newcommand{\ee}{\end{equation}}
\newcommand{\bea}{\begin{eqnarray}}
\newcommand{\eea}{\end{eqnarray}}
\newcommand{\beq}{\begin{equation}}
\newcommand{\eeq}{\end{equation}}
\newcommand{\beqa}{\begin{eqnarray}}
\newcommand{\eeqa}{\end{eqnarray}}
\newcommand{\beqar}{\begin{eqnarray*}}
\newcommand{\eeqar}{\end{eqnarray*}}
\newcommand{\beas}{\begin{eqnarray*}}
\newcommand{\eeas}{\end{eqnarray*}}
\begin{document}

\makeatletter
\renewcommand{\theequation}{\thesection.\arabic{equation}}
\@addtoreset{equation}{section}
\makeatother

\begin{titlepage}

\vfill

\begin{center}
   \baselineskip=16pt
   {\Large\bf  EVH Black Hole Solutions \\ With  Higher Derivative Corrections}
   \vskip 2cm
    Hossein Yavartanoo
       \vskip .6cm
             \begin{small}
      \textit{Department of Physics, Kyung Hee University, Seoul 130-701, Korea}
        \textit{yavar@khu.ac.kr}
        \end{small}\\*[.6cm]
\end{center}

\vfill
\begin{center}
\textbf{Abstract}\end{center}
We analyze the effect of higher derivative corrections to the near horizon geometry of the extremal vanishing horizon (EVH)  black hole solutions in four dimensions.  We restrict ourselves to the Gauss-Bonnet correction with a dilation dependent coupling in an Einstein Maxwell dilaton theory.  This  action may represent the effective action as it arises in tree level heterotic string theory compactified to four dimensions or the K3 compactification of type II string theory. We show that EVH black holes,  in this theory,  develop an AdS$_3$ throat in their near horizon geometry. 
\begin{quote}
\end{quote}
\vfill
\end{titlepage}
\section{Introduction}
It is well known that black holes can be considered as a thermodynamic system with definite temperature and entropy. In order to look into the statistical origin of black hole entropy, we need a quantum theory of gravity. Since string theory gives a framework for studying quantum mechanics of black holes, we carry out our investigation in string theory. 

Recall that for a class of supersymmetric black holes in type IIB string theory, Strominger and Vafa computed the statistical entropy and found agreement with the BekensteinÐHawking entropy \cite{Strominger:1996sh}.  This analysis has been generalized to several other extremal black holes and even non-extremal black holes in different theories of gravity in different dimensions.

During the last several years studying the higher derivative corrections to the entropy of extremal black holes have provided fruitful results in string theory. 
In most of examples, these corrections match the appropriate corrections to the statistical entropy of the corresponding microscopic system.
In many examples in which the identification has been done, black holes possess an AdS$_3$ throat in their near horizon limit and the degeneracy of their microstates can be captured by a two-dimensional conformal field theory (2d-CFT) using AdS$_3$/CFT$_2$ duality.

It has been conjectured \cite{Guica:2008mu} that an extremal Kerr black hole is dual to a chiral 2d-CFT.  The conjecture has been extended to many other extremal black hole solutions in different dimensions \cite{Ext/CFT} (for a recent review of related discussions see  \cite{Simon:2011zza} ).

Although the Kerr/CFT proposal is very interesting, at this stage, its current status is more a suggestion for a possible pair of theories dual to each other and many things should be understood to establish the proposal as a concrete duality.

A precise identification of the proposed chiral CFT is still an open problem and there have been arguments that the Extremal/CFT proposal does not have a dynamical content as the standard AdS/CFT and may only be used for reading the entropy (see   \cite{Kerr-CFT-caveats} for a discussion on this point).  It has been discussed that these AdS$_2$ geometries do not generically represent a decoupled CFT. Even if they would, the AdS/CFT machinery would suggest that these theories are dynamically trivial  \cite{DLCQ-CFT, Sen-AdS2/CFT1}, in the sense that they only contain degeneracy of the vacuum state in their spectrum.

In \cite{SheikhJabbaria:2011gc}, a specific class of extremal black holes, the {\it Extremal Vanishing Horizon} (EVH) black holes is studied. In particular the general four dimensional EVH black
hole solutions have been studied in the Einstein gravity coupled to scalars and gauge fields (for some earlier discussions see e.g. \cite{Fareghbal:2008ar}). This is the general structure of the bosonic part of four-dimensional gauged supergravity theories. 
These supergravity theories themselves could be considered as the low energy effective action of a string compactification to four dimensions. 

For a general stationary black hole solution in these theories it has been shown that the near horizon limit of any EVH black hole, has 
an AdS$_3$ throat. Using this observation the EVH/CFT correspondence has been proposed: gravity on near horizon EVH geometry is described by a 2d CFT.  For some recent discussions see   \cite{deBoer:2011zt}

The purpose of this note is studying the EVH limit in four-dimensional black holes in presence of higher derivative corrections.  These corrections are predicted from string theory.  In this paper 
we shall consider a specific class of higher derivative correction to the action, namely the Gauss-Bonnet term  \cite{Zwiebach:1985uq}. This may arise at tree level in toroidal compactification of heterotic string theory to four dimensions or K3 compactification of type II string theory. 

As we will discuss, near horizon geometry of an EVH black hole in these theories develop an AdS$_3$ throat. This allows us to use the results of EVH/CFT correspondence discussed in \cite{SheikhJabbaria:2011gc}.

%%%%%%%%%%%%%%%%%%%%%%%%%%%%%%%%%%%%%%%%%%%%%%%%%%%%%%%%%%%%%%%%%%%%%%%%%%%%%%%%%%%%%%%%%%%%%%%%%%%
\section{Stationary Black Hole solution with $R^2$ corrections}
%%%%%%%%%%%%%%%%%%%%%%%%%%%%%%%%%%%%%%%%%%%%%%%%%%%%%%%%%%%%%%%%%%%%%%%%%%%%%%%%%%%%%%%%%%%%%%%%%%%%
In this section we study EVH black hole solutions to the heterotic string compactified to four dimensions. Although eventually we shall be interested in studying a general black hole solution
 with all non-zero charges, we shall first consider a black hole solution with a single non-zero scalar and gauge field. The result of our analysis can be simply generalized to the case with more 
scalars and more gauge fields turned on.  In the supergravity approximation the bosonic part of the four dimensional action (in the string frame) is given by
\bea
S=\frac{1}{16\pi G_N} \int d^4x \sqrt{-G}e^{-2\Phi}\bigg(&&\hspace{-7mm}R+4 (\partial \Phi)^2 -F^2 + \mathrm{higher\;derivative\;terms} \cr\cr \label{action1}
&&\hspace{-7mm}+ \mathrm{string\;loop\;corrections} \bigg)
\eea   
In the above action, scalar field is the dilaton and gauge field is the antisymmetric tensor field with one index along the internal manifold.  

Here we consider a specific higher derivative correction to the action, namely  the Gauss-Bonnet term.  At tree level in heterotic string theory this corresponds to an
additional term in the Lagrangian of the following form

\be 
\label{action2}
\Delta{\mathcal L} = \frac{\alpha'}{8}e^{-2\Phi}(R^2-4R_{\mu\nu}R^{\mu\nu} + R_{\mu\nu\alpha\beta} R^{\mu\nu\alpha\beta}).
\ee

Varying the action with respect to metric, scalar and gauge fields  we obtain the following equations of motion:

\bea
\label{EinEq} &&\hspace{-7mm}R_{\mu\nu}-\frac{1}{2}g_{\mu\nu}R = 8\pi G_N\left(T_{\mu\nu}^{\Phi}+  T_{\mu\nu}^{\mathrm{F}}+ T_{\mu\nu}^{\mathrm{G.B}}  \right) \\
&&\hspace{-7mm}\frac{4}{\sqrt{-g}}\partial_{\mu}\left(\sqrt{-g}e^{-2\Phi}g^{\mu\nu}\partial_{\nu}\Phi\right)+\left(R+4(\partial \Phi)^2-F^2\right)e^{-2\Phi}+\Delta {\mathcal L} =0\\
&&\hspace{-7mm}\partial_{\mu}\left(\sqrt{-g}e^{-2\Phi}F^{\mu\nu}\right)=0
\eea
These equations are accompanied by the Bianchi identities for the gauge fields, which is given by 
$F_{[\mu\nu;\lambda]}=0$. The quantities appearing on the right hand side of the Einstein equations (\ref{EinEq}) are defined by
\bea
&&\hspace{-7mm}T_{\mu\nu}^{\mathrm{F}}=\frac{1}{16\pi G_N}\left( 4 F_{\mu\alpha}F_{\nu}^{\;\;\alpha}-g_{\mu\nu}F^2 \right) \\
&&\hspace{-7mm}T_{\mu\nu}^{\Phi}=\frac{1}{4\pi G_N}\left(g_{\mu\nu}\nabla^2\Phi -g_{\mu\nu}(\nabla \Phi)^2-\nabla_{\mu}\nabla_{\nu}\Phi\right) \\
&& \hspace{-7mm}T_{\mu\nu}^{\mathrm{G.B}} =\frac{1}{8\pi G_N}e^{2\Phi}\Delta{\mathcal L}+ \frac{\alpha'}{16\pi G_N}
  \bigg[-2 R_{\mu\alpha}R_{\nu}^{\;\;\alpha} +\frac{1}{2}RR_{\mu\nu}+ \frac{1}{2}R_{\mu\alpha\beta\gamma}R_{\nu}^{\;\;\;\alpha\beta\gamma}\cr\cr
&&
 \hspace{7mm}+e^{2\Phi}\bigg(\frac{1}{2}g_{\mu\nu}\nabla^2(Re^{-2\Phi}) -\frac{1}{2}\nabla_{\mu}\nabla_{\nu}(Re^{-2\Phi})+2\nabla_{\nu}\nabla^{\alpha}(R_{\mu\alpha}e^{-2\Phi})
\cr\cr
&& \hspace{7mm}-\nabla^2(e^{-2\Phi}R_{\mu\nu})
-g_{\mu\nu}\nabla^{\alpha}\nabla^{\beta}(R_{\alpha\beta}e^{-2\Phi})-\nabla^{\alpha}\nabla^{\beta}(R_{\mu\alpha\beta\nu}e^{-2\Phi})\bigg)\bigg]
\eea
It is more convenient to work with the Einstein frame. Transformation to the Einstein frame can be achieved by a conformal rescaling of the metric
\be 
ds_E^2 = e^{-2\Phi} ds_s^2
\ee
Under the above conformal transformation, the Riemannian curvature, the Ricci curvature and the scalar curvature are transformed as (\ref{ReimanT}-\ref{SCT}).
Therefore the transformed action is given by

\be 
S=\frac{1}{16\pi G_N^{(4)}} \int d^4x \sqrt{-G} \left(R -2 (\partial\Phi)^2 - e^{-2\Phi} F^2 + \Delta{\mathcal L} + \alpha' e^{-2\Phi} \Delta{\mathcal L}_{2}  \right)
\ee  
where $\Delta{\mathcal L}_{2}$ denotes the cross terms of the the curvature and $(\partial\Phi)$, which is coming from the Gauss-Bonnet term under the conformal transformation and is given by
\bea 
&&\Delta {\mathcal L}_{2} =
R^{\mu\nu}\nabla_{\mu}\nabla_{\nu}\Phi-R^{\mu\nu}\nabla_{\mu}\Phi\nabla_{\nu}\Phi-\frac{1}{2}R\nabla^2\Phi+(\nabla^2\Phi)^2+(\nabla\Phi)^2(\nabla^2\Phi)\cr\cr
&&\;\;\;\;\;\;-\nabla^{\mu}\nabla^{\nu}\Phi\nabla_{\mu}\nabla_{\nu}\Phi +2 \nabla^{\mu}\Phi\nabla^{\nu}\Phi\nabla_{\mu}\nabla_{\nu}\Phi 
\eea 

A stationary black hole solution (in ADM form) in this theory can be written as

\bea
&& ds^2=-N^2(\rho,\theta)dt^2+ g_{\phi\phi}(\rho,\theta)\left(d\phi+N^{\phi}(\rho,\theta)dt\right)^2+g_{\rho\rho}(\rho,\theta)d\rho^2+g_{\theta\theta}(\rho,\theta) d\theta^2, \cr\cr
&& A=A_t(\rho,\theta)dt+A_{\rho}(\rho,\theta)d\rho+A_{\phi}(\rho,\theta) d\phi, \qquad F=dA \cr\cr
&& \Phi=\Phi(\rho,\theta).
\eea
To write the ansatz for gauge field we have used gauge freedom to set $A_{\theta}=0$.  Let us first rewrite the metric components in following form:
\bea
N^2&=&(\rho-\rho_+)(\rho-\rho_-)\mu(\rho,\theta),\cr\cr
N^{\phi}&=&-\omega+(\rho-\rho_+)\eta(\rho,\theta) ,\cr\cr
g_{\rho\rho}&=&\frac{1}{(\rho-\rho_+)(\rho-\rho_-)\Lambda(\rho,\theta)} , \nonumber
\eea
and we assume that the functions $\mu(\rho,\theta), \Lambda(\rho,\theta)$ do not have zero in $(\rho_+,\infty)$. The black hole temperature is given by
\be
T=\frac{\rho_+-\rho_-}{4\pi}\sqrt{\mu_0(\theta)\Lambda_0(\theta)}
\ee
For a non-extremal  black hole the temperature (surface gravity) is finite and constant on the horizon and so we see that 
$\mu_0(\theta)\Lambda_0(\theta)$ is a constant positive number, which depends on the mass, charges and the angular momentum carrying by black hole. 

Black hole entropy, which is proportional to the area of horizon can be expressed by
\be
S_{BH}=\frac{A_h}{4G_4} = \frac{\pi}{2G_4} \int_0^{\pi} \sqrt{g_{\theta\theta}^{(0)}(\theta)g_{\phi\phi}^{(0)}(\theta)} \;\; d\theta
\ee 
The EVH black hole is defined by limit $S_{BH}\rightarrow 0$ and $T\rightarrow 0$. We may also desire to keep ration $S_{BH}/T$ finite.
Now let us consider the zero entropy limit, $S_{BH}\rightarrow 0$, while we want to keep the geometry regular and non-trivial in this limit. This can be achieved by scaling $g_{\phi\phi}^{(0)} \rightarrow 0$. To get the EVH point and keeping ratio $S_{BH}/T$ finite we want to scale the temperature by $T \sim S_{BH}$. We also want to look at the near horizon of the resulting geometry, means taking the limit $\rho\rightarrow \rho_+$. In this limit the metric is reduced

\be 
ds^2=-\rho \; \tilde{\mu} dt^2 + \frac{d\rho^2}{\rho^2 \tilde{\Lambda}} +\rho\; \tilde{g}_{\phi\phi} \left(d\phi+\tilde{N}^{\phi}dt\right)^2 + \tilde{g}_{\theta\theta}d\theta^2 
\ee 

where $\tilde{\mu}, \tilde{\Lambda}, \tilde{g}_{\phi\phi}, \tilde{N}^{\phi}$ and $\tilde{g}_{\theta\theta}$ are functions of $(\rho, \theta)$ and they are given by taking EVH limit of metric components. In addition $\tilde{\mu}$ and $\tilde{\Lambda}$ does not have any zero in $[0,\infty)$. Using the fact that the temperature (surface gravity) must be independent of the angular coordinate, one can conclude that $\tilde{\mu}(\rho,\theta)\tilde{\Lambda}(\rho,\theta)$ is a positive constant ($\theta$ independent) at $\rho=0$.

To get the near horizon geometry we need to scale radial coordinate $\rho= \epsilon r^2$ and taking $\epsilon\rightarrow 0$. The resulting geometry after rescaling \cite{SheikhJabbaria:2011gc} is given by

\be\label{AdS3-solution}
ds^2= a({\theta}) \left(-r^2 d\tilde{t}^2 + L^2  \frac{dr^2}{r^2} + b({\theta}) r^2 d{\phi}^2 + R^2d{\theta}^2\right)
\ee

The above metric is specified by two parameters $R$ and $L$.  $R$, as long as periodicity of ${\theta}$ direction is not specified, may be absorbed in
the definition of  ${\theta}$, while $L$ is the radius of AdS$_2$ part of the metric. 

Besides the metric we also have gauge and scalar fields which we need to analyze their behavior in EVH limit.  We assume that in a generic point at the near horizon, the scalar field remains finite and follows the symmetries of the metric.  Therefore at the near horizon it is a function of only angular coordinate ${\theta}$.  Gauge fields should also follow the symmetry of the metric (\ref{AdS3-solution}).  

If we plug this ansatz into the equations of motion, we obtain a system of coupled differential equations on functions appearing in the metric, gauge and scalar fields.  When $\alpha'=0$,  we can solve the Einstein, gauge field and scalar field equations explicitly \cite{SheikhJabbaria:2011gc} . After rescaling the coordinate, the solutions can be expressed as follows:
\bea
\label{solution0}&& ds^2 = R_{AdS3}^2 |\sin\theta | \left(-r^2 d{\tau}^2 +   \frac{dr^2}{r^2} +  r^2 d{\psi}^2 + \frac14 d{\theta}^2\right)\,,\\
&& F_{\mu\nu}=0\,,\;\;\; e^{\frac{2\Phi^{(0)}}{\sqrt3}}= g_0 \tan \frac{\theta}{2}\,
\eea

For non-zero $\alpha'$, the expressions for equations of motion are too lengthy to be displayed here. In addition these equations are too complicated  to be solved analytically. 

\section{Non-perturbative solution}
  
  Although we are more interested in solutions which are smoothly connect to solution (\ref{solution0}), but efore analyzing the general solution, we observe that equations of motion admit the following simple solutions:
  \be 
  b(\theta)=b_0,\quad a(\theta)=\frac{\alpha'}{4L^2x_0^6\sin^2\frac{R\theta}{L}} \;, \quad e^{\Phi}=x_0, \qquad  0<\alpha'<4,
  \ee
where $b_0$ is an arbitrary constant  and $x_0$ is the positive root of the following equation
\be
40x_0^ 4+8x_0^2-4+\alpha'=0
\ee
This geometry is smooth everywhere and singular points in the metric are a coordinate singularity.   The solution vanishes when $\alpha'$ goes to zero, therefore it does not connect smoothly to solution (\ref{solution0}).  

\section{Perturbative analysis}

In order to analyze the system of coupled differential equations perturbatively in $\alpha'$ parameter, let us assume all functions appearing in the solution are analytical functions of parameters $\alpha'$.  A series solution to the equations of motion can be developed simply by expanding
\be
X=X^{(0)} + \alpha' X^{(1)} + \cdots
\ee
where $X$  represent any of metric components, gauge field and scale fields.  It is important to notice that action (\ref{action1}) with (\ref{action2}) is computed up to the first order of $\alpha'$ itself and we have ignored all other higher order  terms in $\alpha'$ expansion of the effective action.

After expanding all functions and plugging these into the equations of motion we get a series of differential equations and one can try to solve these equations order by order to find a perturbative solution in $\alpha'$ expansion.

\begin{description}
  \item[Zeroth order:] At zeroth order in $\alpha'$ expansion the gauge field equations  along with Einstein equations reduces to the following equations: 
\bea
 \frac{d b^{(0)}}{d \tilde\theta}=0 \qquad & \Rightarrow &\qquad b^{(0)}=b_0=const.\ ,\\  \nonumber
\frac{d^2 a^{(0)}}{d\tilde\theta^2}+\frac{4R^2}{L^2}a^{(0)}  = 0  &\Rightarrow& \qquad a^{(0)} =a_0\sin\frac{2R}{L}\tilde\theta\,,\nonumber
\eea

The $\theta\theta$-component of the Einstein equations, which is compatible with the scalar field $\Phi$ equation of motion, yields
\be
\label{solphi}
\frac{d\Phi^{(0)}}{d\tilde{\theta}}= \pm \frac{\sqrt3 R}{L} \frac{1}{\sin(\frac{2R}{L}\tilde{\theta})}\ \qquad \Longrightarrow \qquad  e^{\frac{2\Phi^{(0)}}{\sqrt3}}= g_0 \tan \frac{R\tilde\theta}{L}\,,
\ee
where $g_0$ is a constant.  Therefore the final solution to the zeroth order equations, after a rescaling,  is specified by only two parameters, $g_0$ and the AdS$_3$ radius $R_{AdS3}$.  The solution can be written as

\bea
&& ds_{(0)}^2 = R_{AdS3}^2 |\sin\theta| \left(-r^2 d{\tau}^2 +   \frac{dr^2}{r^2} +  r^2 d{\psi}^2 + \frac14 d{\theta}^2\right)\,,\\
&& F^{(0)}_{\mu\nu}=0\,,\;\;\; e^{\frac{2\Phi^{(0)}}{\sqrt3}}= g_0 \tan \frac{\theta}{2}\,
\eea
where $\theta\in [0,\pi]$ and $\psi\in [0,2\pi\epsilon]$. We stress that the AdS$_3$ throat which has developed in the near horizon limit of the EVH black hole  is a {\it pinching} AdS$_3$, because for the circle inside
AdS$_3$, $\psi$  has a vanishing periodicity  $2\pi{\epsilon}$.

\end{description}

\begin{description}
  \item[First order:] 
   At the first order in $\alpha'$ expansion the $r\theta$-component of the Einstein equations along with gauge field equations   imply
  \be
  F^{(1)}_{\mu\nu}=0,\qquad \frac{db^{(1)}}{d\theta}=0.
  \ee
  There are two other independent components of the Einstein equations, namely $tt$ and $\theta\theta$ components, which should be solved along with the scalar field equation.  These equations determine functions $a^{(1)}$ and $\Phi^{(1)}$.  The explicit forms of these equations are too lengthy to be written here, but one can find that an appropriate combination of Einstein equations gives us a differential equation on $a^{(1)}$.  Although this equation is too complicated to be solved analytically but one can try to solve in form of a power series expansion near any arbitrary point. In summary we find that the field configuration at the first order in $\alpha'$ expansion after a rescaling is given by
  \bea\label{NHG}
 && ds^2 = a(\theta) \left(-r^2 d{\tau}^2 +   \frac{dr^2}{r^2} +  r^2 d{\psi}^2 + \frac14 d{\theta}^2\right)\,,  \\
&& F_{\mu\nu}=0\,,\qquad \Phi=\Phi(\theta)\, \nonumber
  \eea   
 We obtained again a geometry which is conformal to an AdS$_3$ times an interval.  The AdS$_3$ throat is a pinching AdS$_3$, as we have discussed. 
  \end{description}

  \begin{description}
  \item[Higher order results:] 
One can continue the perturbative analysis  to the higher orders in $\alpha'$.  Equations of motion are getting more complicated in higher order of perturbation.  But it is rather straightforward to check that  $r\theta$-component of the Einstein equations and gauge field equations at each order can be solved to find that the gauge fields are vanishing and the function $b$ remains a constant . Indeed it is not difficult to check that $F=0$ and $b=const.$ solve the complete equations of motion. Two other independent components of the Einstein equations and scalar field equation are left to be solved for two functions $\Phi(\theta)$ and $a(\theta)$. Although we cannot solve these equations analytically but near any given point $\theta_0$, we can solve $a$ and $\phi$ as power series in $\theta$. This completes our proof that (\ref{NHG}) is a solution to the equation of motion.

\end{description}
\section{Conclusion}
In this note we have studied near horizon geometry of an EVH black hole in presence of higher derivative correction.  We have focused on a special class of $R^2$ corrections i. e.  Gauss-Bonnet term.  In four dimensions the Gauss-Bonnet term is a topological invariant and it does not contribute to the Einstein equations, but it can contribute to the field equations if it couples to a scalar field. 
This is the case we have for instance in compactification of the heterotic string theory where we have a mixing between Gauss-Bonnet term and dilaton. It is known that the Gauss-Bonnet terms appears at order $\alpha'$ in the heterotic string effective action \cite{Zwiebach:1985uq}.  

In \cite{SheikhJabbaria:2011gc}, it has been shown that the near horizon limit of any EVH black holes on the Einstein theory in four dimensions has an AdS$_3$ throat. Using the AdS/CFT correspondence we can study the conformal field theory dual to this near horizon geometry. 
Indeed it has been argued that gravity on the near horizon geometry of an EVH black hole is dual to, or described by a two dimensional conformal field theory. It also has been shown that this two-dimensional conformal field theory connects to the chiral conformal field theory in the  Kerr/CFT correspondence. It has been proposed that the EVH/CFT in the DLCQ description reproduces Kerr/CFT.

In this work we have studied the effect of higher derivative correction to this picture. To be more precise, we have considered a general rotating black hole solution in Einstein gravity coupled with a scalar and a gauge field along with a $R^2$ correction in Gauss-Bonnet form. We have considered that the scalar field couples to the Gauss-Bonnet term. In this case although the Gauss-Bonnet term is a topological invariant in four dimensions but because of its coupling to the scalar field it contributes to the equation of motion. In this paper we have analyzed the effect of this correction to the near horizon geometry of the EVH black hole. The result of this work can be simply generalized to the case with more than one gauge field and scalar field. Indeed this is the situation we have in toroidal compactification of heterotic string theory to four dimensions.

In this work we have shown that the near horizon geometry of an EVH black hole with the Gauss-Bonnet correction term contains an AdS$_3$ throat just like the case without higher derivative corrections. The Gauss-Bonnet correction modifies the scalar field behavior and the pre-factor $a(\theta)$ in the near horizon geometry. Having an AdS$_3$ throat, we can use AdS/CFT to study the field theory dual to this near horizon along the same line as discussed in \cite{SheikhJabbaria:2011gc}.

\section*{Acknowledgments}
I would like to thank  Eoin \'O Colg\'ain and Shahin Sheikh-Jabbari for useful comments and discussion. I also would like to thank Raju Roychowdhury for his collaboration at the early stage. This work was supported by the National Research Foundation of Korea Grant funded by the Korean Government (NRF-2011-0023230).

%%%%%%%%%%%%%%%%%%%%%%%%%%%%%%%%%%%%%%%%%%%
%%%%%%%%%%%%%%%%%%%%%%%%%%%%%%%%%%%%%%%%%%%
\appendix
%%%%%%%%%%%%%%%%%%%%%%%%%%%%%%%%%%%%%%%%%%%
%%%%%%%%%%%%%%%%%%%%%%%%%%%%%%%%%%%%%%%%%%%

\section{Conformal transformation}
The action (\ref{action1}) does not have the familiar form of the Einstein-Hilbert term because of the
factor of $e^{-2\Phi}$ in front of the scalar curvature.  This factor reflects the fact that the action has been computed in string perturbation theory (string frame) and such terms typically scale as $1/g_s^2$ where $g_s$ is string coupling constant.
To write the action in a more familiar form, we can make the following redefinition of the metric:
\be
g_{\mu\nu}^{(s)} \rightarrow g_{\mu\nu}^{E}=e^{-2\Phi} g_{\mu\nu}^{(s)}.
\ee

Under this transformation, the Riemann tensor, the Ricci tensor and the Ricci scalar are transformed as follows:

\bea 
\label{ReimanT}
\hspace{-5mm}R_{\mu\nu\alpha\beta}&\rightarrow&e^{2\Phi}\bigg[R_{\mu\nu\alpha\beta} -g_{\mu\beta}\triangle_{\alpha\nu} \Phi   -g_{\nu\alpha}\triangle_{\mu\beta} \Phi +g_{\nu\beta}\triangle_{\mu\alpha} \Phi+ g_{\mu\alpha}\triangle_{\nu\beta} \Phi\cr\cr &&\;\;\;\;\;\;+(g_{\mu\beta}g_{\nu\alpha}-g_{\mu\alpha}g_{\nu\beta})(\nabla\Phi)^2     \bigg]\\ 
\hspace{-5mm}\label{RicciT}R_{\mu\nu}&\rightarrow& R_{\mu\nu} -2\nabla_{\mu}\nabla_{\nu}\Phi-g_{\mu\nu}\nabla_{\sigma}\nabla^{\sigma}\Phi+2\nabla_{\mu}\Phi\nabla_{\nu}\Phi-2g_{\mu\nu}( \nabla \Phi)^2 \\
\hspace{-5mm}\label{SCT}R&\rightarrow& e^{-2\Phi}\left[ R - 6 (\partial \Phi)^2 - 6 \nabla^2 \Phi \right]
\eea                                                               
where 
\be
\triangle_{\mu\nu} \Phi =\nabla_{\mu}\Phi\nabla_{\nu}\Phi-\nabla_{\mu}\nabla_{\nu}\Phi   
\ee
In the above expressions all derivatives and metric appear on the right side are in an Einstein frame. 
\section{Variation Of The Riemann Tensor, The Ricci Tensor, And The Ricci Scalar}
To calculate variation of the Riemann curvature, Ricci tensor and Ricci scalar we start with the definition of the Riemann curvature tensor which is given by
\be
 R^{\rho}_{\sigma\alpha\beta}=\partial_{\alpha}\Gamma^{\rho}_{\beta\sigma}-\partial_{\beta}\Gamma^{\rho}_{\alpha\sigma} + \Gamma^{\rho}_{\alpha\lambda}\Gamma^{\lambda}_{\beta\sigma}-
 \Gamma^{\rho}_{\beta\lambda}\Gamma^{\lambda}_{\alpha\sigma}.
\ee
The Riemann curvature depends only on the Levi-Civita connection, therefore the variation of the Riemann tensor can be expressed as
\be
 \delta R^{\rho}_{\sigma\alpha\beta}=\partial_{\mu}\delta\Gamma^{\rho}_{\beta\sigma}-\partial_{\beta}\delta\Gamma^{\rho}_{\alpha\sigma} + \delta \Gamma^{\rho}_{\alpha\lambda}\Gamma^{\lambda}_{\beta\sigma}+\Gamma^{\rho}_{\alpha\lambda}\delta\Gamma^{\lambda}_{\beta\sigma}-
 \delta\Gamma^{\rho}_{\beta\lambda}\Gamma^{\lambda}_{\alpha\sigma} -
 \Gamma^{\rho}_{\beta\lambda}\delta\Gamma^{\lambda}_{\alpha\sigma}.
 \ee
Since we have the difference of two connections, it is a tensor and we can thus calculate its covariant derivative and then it is easy to observe that the expression for the variation of Riemann curvature tensor above is equal to the difference of two covariant derivatives,
\be
\label{delRieman}
 \delta R^{\rho}_{\sigma\alpha\beta}=\nabla_{\alpha}\left(\delta \Gamma^{\rho}_{\beta\sigma} \right)-\nabla_{\beta}\left(\delta \Gamma^{\rho}_{\alpha\sigma} \right)
\ee
Now, using the fact that the difference of two connections, $\delta\Gamma^{\rho}_{\nu\sigma} $  should transform as a tensor, it can be written as
\be
\label{deltachr}
\delta \Gamma^{\rho}_{\beta\sigma}= \frac{1}{2}g^{\rho\lambda}\left(\nabla_{\beta} \delta g_{\sigma\lambda}    +\nabla_{\sigma}\delta g_{\beta\lambda} -\nabla_{\lambda}\delta g_{\sigma\nu}   \right)
\ee
Combining it with equation (\ref{delRieman}) we get the following expression for the variation of the Riemann curvature tensor in terms of the variation of the metric:
\bea 
\delta R_{\rho\sigma \alpha\beta} = \bigg( && \hspace{-6mm}-R_{\nu\sigma\alpha\beta} g_{\rho\mu} -\frac{1}{2} g_{\rho\mu}g_{\beta\nu}\nabla_{\sigma\alpha}  -\frac{1}{2} g_{\sigma\mu}g_{\alpha\nu}\nabla_{\rho}\nabla_{\beta} + \frac{1}{2} g_{\rho\mu}g_{\alpha\nu}\nabla_{\sigma}\nabla_{\beta}  \cr\cr 
&&\hspace{-6mm}+ \frac{1}{2} g_{\sigma\mu}g_{\beta\nu}\nabla_{\rho}\nabla_{\alpha} \bigg)\delta g^{\mu\nu}
\eea 
We may now obtain the variation of the Ricci curvature tensor simply by contracting two indices of the variation of the Riemann tensor,
\be
 \delta R_{\alpha\beta}\equiv \delta R^{\rho}_{\alpha\rho\beta}=\nabla_{\rho}\left(\delta \Gamma^{\rho}_{\beta\alpha} \right)-\nabla_{\beta}\left(\delta \Gamma^{\rho}_{\rho\alpha} \right)
\ee 
This can be expressed in terms of the metric variation as follows
\be 
\delta R_{\alpha\beta}=\left(\frac{1}{2}g_{\alpha\mu}g_{\beta\nu}\nabla^2 + \frac{1}{2}g_{\mu\nu}\nabla_{\alpha}\nabla_{\beta}-\frac{1}{2}g_{\beta\mu}\nabla_{\alpha}\nabla_{\nu}- \frac{1}{2}g_{\alpha\mu}\nabla_{\beta}\nabla_{\nu} \right)\delta g^{\mu\nu}
\ee 

The Ricci scalar is defined as 
\be
R\equiv g^{\mu\nu} R_{\mu\nu}.
\ee
Therefore, its variation is given by
\be
\delta R= R_{\mu\nu}\delta g^{\mu\nu} + g^{\mu\nu}\nabla_{\rho}\left(\delta \Gamma^{\rho}_{\nu\mu} \right)-g^{\mu\nu}\nabla_{\nu}\left(\delta \Gamma^{\rho}_{\rho\mu} \right)
\ee 
and substituting in the equation  (\ref{deltachr}) one finds
\be
\delta R= \left(R_{\mu\nu} +g_{\mu\nu}\nabla^2 -\nabla_{\mu}\nabla_{\nu} \right)\delta g^{\mu\nu}
\ee

\end{document}